\documentclass[a4paper,11pt]{article}
\usepackage[a4paper,margin=2.0cm]{geometry}
\usepackage{authblk}
\usepackage{lineno}
\usepackage{graphicx}
\newcommand{\hls}{\textsf{hls4ml}}
\usepackage[inline]{enumitem}
\usepackage{paralist}
\usepackage{url}
\usepackage{hyperref}

\title{\boldmath Machine Learning-Based Reconstruction for Resistive Silicon Sensors}
\author[1]{Alexander Aoki,}
\author[1]{Gaetano Barone\thanks{Corresponding author.},}
\author[2]{Leena Diehl,}
\author[3]{Gabriele Giacomini,}
\author[2]{Vagelis Gkougkousis,}
\author[1]{Hanshal Goyal,}
\author[1]{Rohan Kher,}
\author[4]{Daniel Li,}
\author[2]{Anna Macchiolo,}
\author[2]{Yevhenii Padnuik,}
\author[2]{Daria Senina,}
\author[1]{Samantha Sunnarborg,}
\author[1]{Jessica Tang,}
\author[3]{Alessandro Tricoli,}
\author[1]{Lixing Wang}
\author[1]{and Don C. Wong}

\affil[1]{Brown University, 182 Hope Street, Providence, RI 02912, USA}
\affil[2]{University of Zurich}
\affil[3]{Brookhaven National Laboratory}
\affil[4]{Northwestern University}

\begin{document}
\maketitle

\abstract{
Low-Gain Avalanche Diodes (LGADs) and AC-coupled Low-Gain Avalanche Diodes (AC-LGADs) are promising technologies for precision timing and four-dimensional
tracking. In AC-LGADs, the AC pad is coupled to the resistive n$^{+}$ layer through a dielectric layer, while the gain layer remains unsegmented. This structure provides a 100\% fill factor and enables good spatial resolution with a relaxed readout pitch. The same signal-sharing mechanism that makes interpolation possible complicates the readout: charge spreads across multiple pads, the useful information can approach the electronic-noise threshold, and matrix-inversion approaches can become computationally challenging and sensitive to off-diagonal noise. In this work, we study machine-learning-based reconstruction and compression for resistive silicon sensors. We use full-waveform information from correlated pads to regularise the reconstruction and extract spatial information beyond what is available from binary readouts or reduced-amplitude summaries. We first introduce recurrent neural network models based on LSTM layers, which provide a proof-of-concept implementation for full-waveform reconstruction and have been tested for FPGA deployment using \hls. We also study routes towards bandwidth reduction with waveform rasterisation and window-selection methods, and extend the approach beyond the first model to topology-agnostic transformer-based architectures that use pad coordinates as part of the input. These models are designed to support arbitrary pad counts and geometries, mitigate edge distortions, preserve approximately $10~\mu\mathrm{m}$ position resolution for $500~\mu\mathrm{m}\times500~\mu\mathrm{m}$ pitched sensors, and guide future resistive-silicon sensor designs.}

%\keywords{Performance of High Energy Physics Detectors, Software architectures (event data models, frameworks and databases), Analysis and statistical methods, Data processing methods, Data reduction methods, Electronic detector readout concepts (solid-state), Hybrid detectors, Timing detectors, Radiation-hard detectors, Si microstrip and pad detectors Solid state detectors}

%\flushbottom

\section{Introduction}
\label{sec:introduction}
Precision timing and spatial resolution are indispensable for next-generation experiments such as the HL-LHC~\cite{CMS_TDR_MTD,CMSCollaboration_2008}, future $e^+e^-$ colliders, and the EIC, where 4D tracking suppresses pile-up, enables time-of-flight PID, and reconstructs dense vertices; similar needs arise in fixed-target~\cite{ROYON2024168886} and space missions~\cite{GAUTIER2021165599,WU20192672}, and are highlighted in community roadmaps~\cite{ECFA_Roadmap_2021}.  Low-Gain Avalanche Diodes (LGADs) and AC-coupled Low-Gain Avalanche Diodes (AC-LGADs) are silicon sensors with internal gain that provide precision timing. LGADs meet timing requirements with ${\cal O}(20$--$50)$~ps performance~\cite{Pellegrini2014,Wu_2020,cartiglia2020,investigationchargesharing,lgad1,lgad2}, yet typical $\sim1\times1$~mm$^2$ pads limit spatial resolution. In the AC-LGAD variant~\cite{Mandurrino2020_NIMA959,tornago2021_nima1003_165319}, the AC pad is coupled to the resistive n$^{+}$ layer through a dielectric layer. The gain layer is not segmented, resulting in a 100\% fill factor. This allows the sensor to preserve the timing performance associated with LGAD technology while using charge sharing to obtain spatial resolution with a readout pitch that can be more relaxed than the target spatial precision. We target high-energy physics applications in low-to-moderate radiation regimes, including the FCC-ee and upgrades to LHC experiments. For the High-Luminosity LHC, possible applications include the LHCb VELO upgrade and a CMS tracker Phase-3 upgrade, where AC-LGADs could extend timing capabilities in the forward region, increase rapidity coverage, or replace one or two disks with timing-capable instrumentation. For FCC-ee, timing in the outermost silicon layers could enhance particle identification and reduce the systematic uncertainty on beam energy. Additional applications include time-of-flight systems and medical detectors.

Sensor fabrication can be targeted to the needs of a specific application by tuning the sensor size, gain characteristics, and depletion-layer size. Application-specific design can also address constraints on material budget, energy consumption, and climate-change resilience. The readout and the sensor design must therefore be considered together, since the achievable position reconstruction and timing performance depend on the details of the charge collection and the readout geometry.

\section{Charge sharing in resistive silicon sensors}
\label{sec:charge-sharing}

Increased charge sharing is an intrinsic property of RSDs and AC-LGADs. When a particle interacts with the sensor, the induced signal is not confined to a
single readout pad. Instead, multiple pads receive correlated waveforms. This behaviour is useful because it allows interpolation within a readout cell, but it also turns position reconstruction into a coupled multi-channel problem.

For a pad $i$, with $\alpha_i$ representing the area of the pad as viewed from the true hit position and $r_i$ representing the distance between the true hit and that pad, we parameterize the signal fraction as~\cite{tornago2021_nima1003_165319}
\begin{equation}
\label{eq:charge-sharing}
S_i =
\frac{\alpha_i/\ln r_i}{\sum_i^n \alpha_i/\ln r_i} \, .
\end{equation}
\noindent Different pad arrangements therefore produce different signal-sharing patterns and, consequently, different reconstruction performance. Hits well inside a pad have amplitudes that are independent of position. At larger distances, the amplitude depends strongly on position due to charge sharing through the sensor's inter-pad resistance characteristics. As the distance from the pad centres increases, the signal approaches the noise-dominated regime~\cite{apresyan2020,heller2022_aclgad_bnl_hpk_jinst,DAmen222}. The main question this work is beginning to address is how much information near the noise threshold can be recovered and used for position reconstruction. The maximum information about the position can be obtained by inverting the waveform matrix as a function of pad geometry, relating the true hit position to the signal collected by each electrode. This approach becomes challenging when the matrices are ill-conditioned. Large off-diagonal elements, introduced by signal sharing beyond immediate neighbours, and electronic noise can produce large fluctuations and biases. Centroid and other charge-imbalance methods can be viewed as simplified matrix-inversion procedures: they regularise these fluctuations but use less information, thereby degrading the achievable precision relative to the full potential of the waveform data. We therefore introduce machine learning to regularise the reconstruction and extract maximal information from the correlated waveforms. 

Early ML efforts use pulse height or time-over-threshold as features, improving robustness but still missing correlations between waveform shape, rise time, and dependence on Landau fluctuations. Undoubtedly, more information is to be extracted from the sensors, pushing the noise limit.  More recent work has demonstrated significant progress: \begin{inparaenum}[(i)] \item dense neural networks achieve 2--10~$\mu$m at 100--200~$\mu$m pitch~\cite{siviero2021,investigationchargesharing}, \item random forests reach $\sim 8~\mu$m at 200~$\mu$m pitch~\cite{tornago2023}, \item neural regression in beam tests delivers 5--10~$\mu$m~\cite{heller2022_aclgad_bnl_hpk_jinst}, and \item previous work on neural networks trained full waveforms processing and regressions show a preliminary performance of  $\sim 10~\mu$m at $500 \times 500~\mu$m pitch~\cite{baronedrd3,Barone2025}. \end{inparaenum}  We report updates on the preliminary results, and we organise the  work around two related efforts. The first is the integration of high-throughput machine-learning-based readout. The second is the study of compression-throughput methods for complex sensor geometries for devices with $p$-type substrates with varying geometrical pad arrangements. Both rely on the same hypothesis: the full waveform carries information that is not captured by binary readout or by a small number of amplitude-derived features.

\section{Sensors and experimental setup}
\label{sec:sensors-setup}

\begin{table}[htbp]
\label{tab:bnl-sensors}
\centering
\caption{BNL sensor information used in this study.}
\smallskip
\begin{tabular}{ll}
\hline
Item & Information\\
\hline
Producer & Brookhaven National Laboratory\\
Implant type & Uniform p-type implants on the silicon surface\\
Active area & $1.3~\mathrm{mm}^{2}$\\
Gain-layer dose, $30~\mu\mathrm{m}$ sensors & $2.7\times10^{12}~\mathrm{cm}^{-2}$\\
Gain-layer dose, $20~\mu\mathrm{m}$ sensors & $2.25\times10^{12}~\mathrm{cm}^{-2}$\\
Baseline geometry & Square grid\\
Secondary geometries & Diamond, triangular, and other geometries\\
\hline
\end{tabular}
\end{table}

\noindent We study BNL-produced sensors with uniform p-type implants on the silicon surface. We consider devices with an active area of $1.3~\mathrm{mm}^{2}$ and
gain-layer doses of $2.7\times10^{12}~\mathrm{cm}^{-2}$ for $30~\mu\mathrm{m}$ sensors and $2.25\times10^{12}~\mathrm{cm}^{-2}$ for
$20~\mu\mathrm{m}$ sensors. The baseline arrangements are square grids, with secondary arrangements including diamond, triangular, and other geometries. Table~\ref{tab:bnl-sensors} summarises the sensors we considered.

\begin{figure}[htbp]
\centering
\includegraphics[width=.5\textwidth]{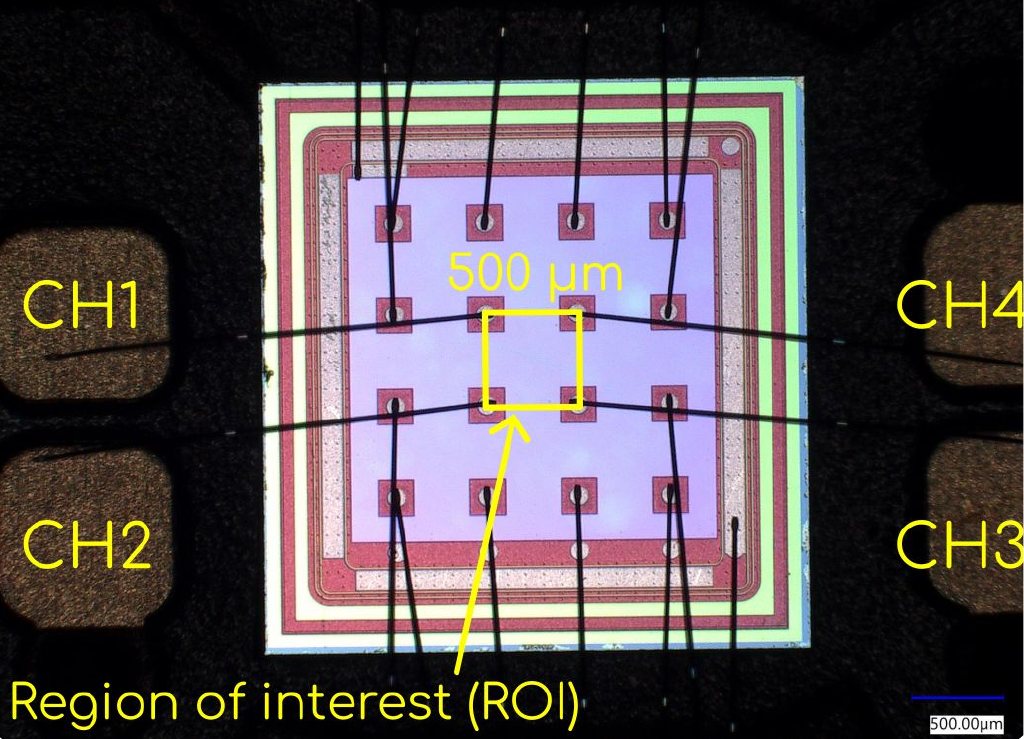}

\caption{BNL AC-LGAD sensor and readout layout, including the four-channel readout, the $500~\mu\mathrm{m}$ region of interest, and the AC-LGAD device summary.\label{fig:sensor-setup}}
\end{figure}

The sensors considered here are unirradiated and have a $200~\mu\mathrm{m}$ pad size, a $500~\mu\mathrm{m}$ pitch, and a $2\times2$ pad readout, with unused pads connected to ground, as shown in Figure~\ref{fig:sensor-setup}. For readout, we use the Chubut2 readout board~\cite{chubut} readout out by a CAEN DT5742 digitiser, 500 MHz with 5 GS/s. The transient current technique (TCT) data were collected at the setup of the University of Zurich with a 1064 nm laser with a spot Gaussian size of 9 $\mu$m, with intensity calibrated to match 1 MIP, and a resolution of 1 $\mu$m scanning 385 grid points with about 100 events per point. The sensors are biased at full depletion voltage.

\section{Waveform machine-learning-based reconstruction}
\label{sec:ml-strategy}

As introduced, we study how machine learning can regularise charge-sharing reconstruction and extract maximal information from the measured waveforms. First, an approach not tied to a fixed pad arrangement can help identify geometries that maximise positional resolution. Second, full waveform processing can use shape
correlations between the leading pad and neighbouring pads. Third, exploiting all pads simultaneously allows the network to use cross-channel correlations that are otherwise lost in simplified readout schemes.

Although the main long-term challenges to consider are Landau fluctuations, performance degradation due to radiation damage, and limitations imposed by shaping, ASIC/readout, and electronics constraints, this study is meant to demonstrate the potential of these methods. To address Landau fluctuations, we aim to introduce laser-assisted test-beam training and parameterise the response as a function of deposited charge. Radiation-damage effects must eventually be parameterised as a function of fluence. To address electronics constraints, the longer-term direction is to implement the processing in off-detector electronics or on an FPGA.

\begin{figure}[htbp]
\includegraphics[width=0.5\textwidth]{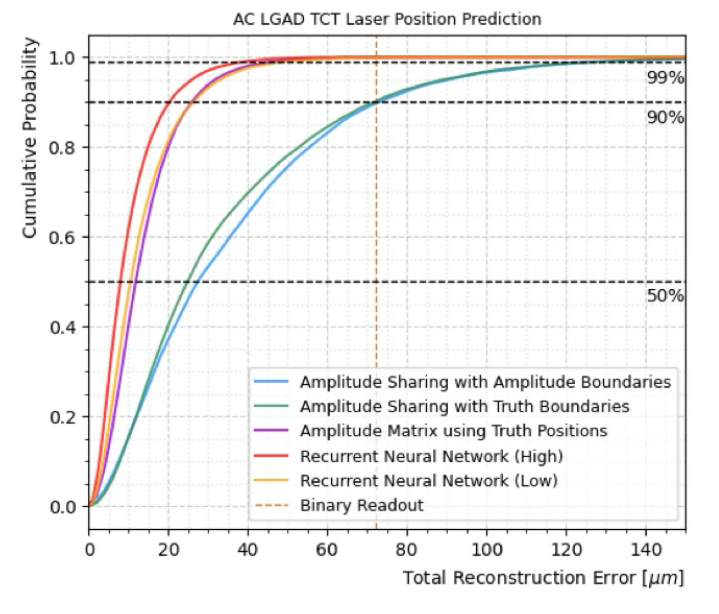}
\includegraphics[width=0.5\textwidth]{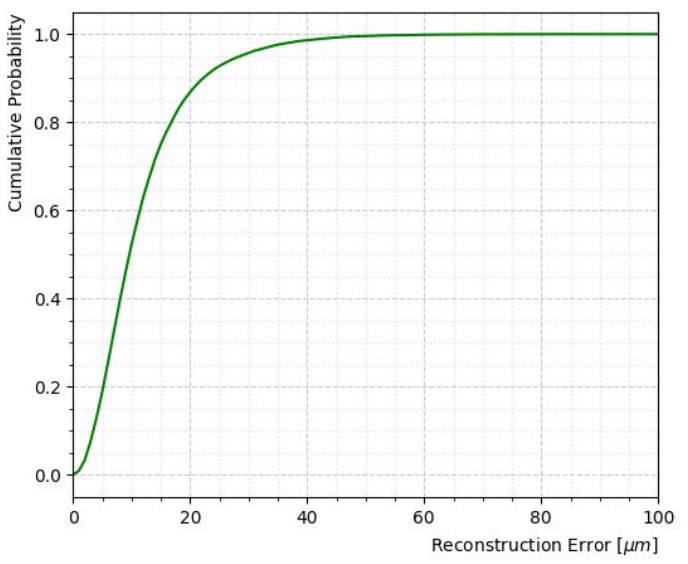}

\caption{The figures show the cumulative probability of reconstructing a given hit position as a function of its total reconstruction resolution. The figure on the left compares analytical matrix-inversion methods based on amplitude and full-matrix inversion methods to the RNN-readout model. The figure on the right evaluates the performance of the RNN-based reconstruction algorithm when simulated with \hls.
\label{fig:ml-models}}
\end{figure}

\begin{figure}[htbp]
\centering
\includegraphics[width=0.8\textwidth]{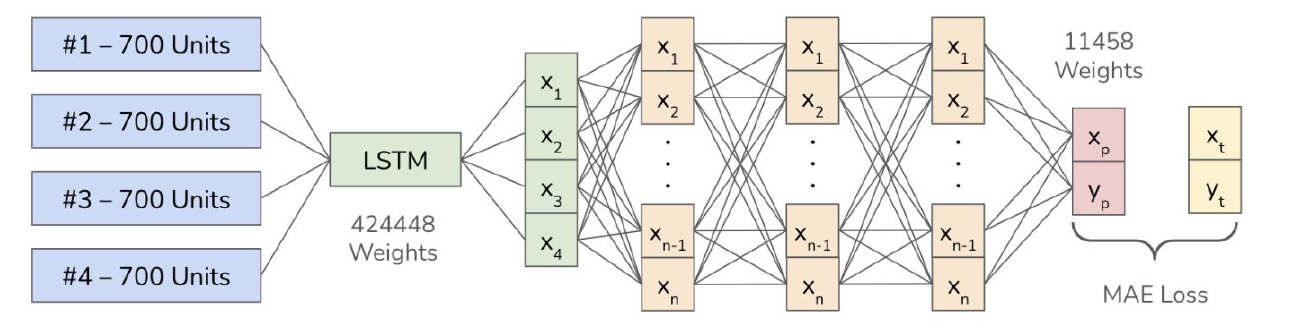}

\caption{Structure of the RNN network with Long Short-Term Memory layer. \label{fig:rsd-readout}}
\end{figure}

\noindent However, our intended readout chain performs data reduction toward off-detector electronics while preserving the position information contained in the
correlated pad waveforms. The first full-waveform reconstruction model uses a recurrent neural network with Long Short-Term Memory (LSTM) layers, as depicted in Figure~\ref{fig:rsd-readout}. The input consists of complete waveforms from TCT scans, and the network is trained to predict the hit position. With this model, and as shown in Fig.~\ref{fig:ml-models}, we achieve a 50\% improvement over matrix-inversion methods and a seven-fold improvement over binary readout, indicating a potential resolution of about $10~\mu\mathrm{m}$ from pixels with $500~\mu\mathrm{m}\times500~\mu\mathrm{m}$ pitch. We also translated the initial model with \hls~\cite{fastml_hls4ml, fahim:2021cic, Duarte:2018ite, schulte2025hls4mlflexibleopensourceplatform} to an FPGA-compatible architecture as a proof of concept and to study the impact of information loss.

\begin{figure}[htbp]
\centering
\includegraphics[width=0.4\textwidth]{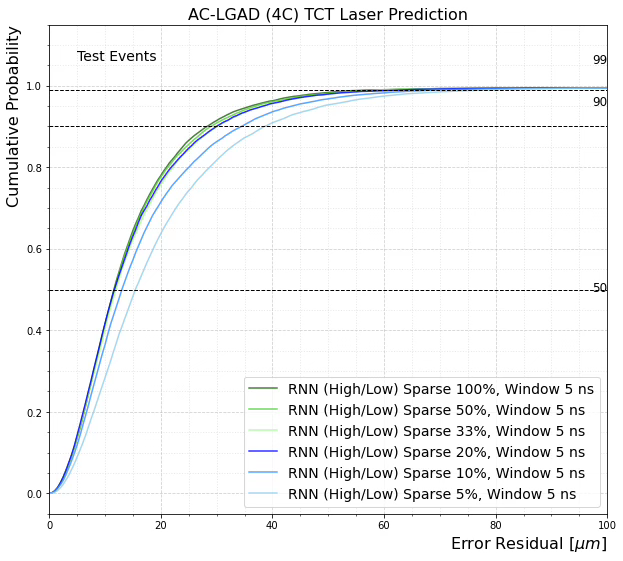}
\centering
\includegraphics[width=0.4\textwidth]{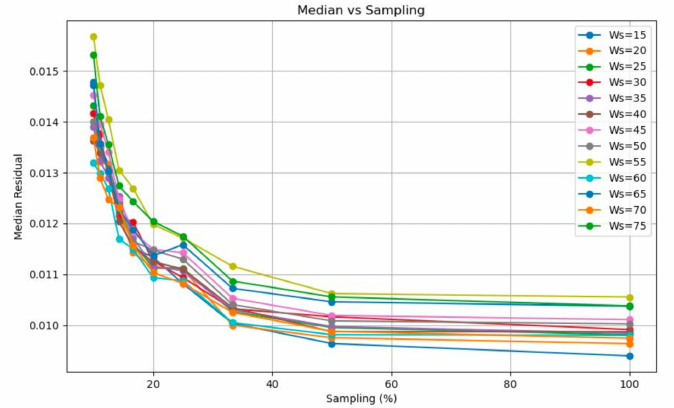}

\caption{On the left, the cumulative probability of the reconstruction error is shown for different sparsity levels for a fixed window size. On the right, the median reconstruction uncertainty is shown as a function of the sampling rate for different window sizes in the number of digitiser samples. \label{fig:rasterisation}}
\end{figure}

The full waveform contains spatial information, but preserving every sample from
every pad can be costly in bandwidth and latency. We, therefore, study compression and information loss using rasterisation. With approximately 70\% compression, we observe less than 5\% loss in resolution. For each window size, sparsity is varied and performance is tested. We observe a performance divergence for all window sizes below a sparsity of about
20\%. Window selection is used to preserve the relevant part of the waveform. For a given sparsity, the RNN performance does not change significantly as long as the peak is contained inside the selected window, as shown in Fig.~\ref{fig:rasterisation}. Even if waveform rasterisation in training and evaluation remains preliminary, the current studies are compatible with $10$--$15~\mu\mathrm{m}$ resolution on $500~\mu\mathrm{m}\times500~\mu\mathrm{m}$ pixels.

Our first model has limitations because its geometry is fixed within the architecture's spatial encoding. It also depends strongly on a fine-grained
training grid that maps each point response to a set of $n$ waveforms or responses per hit. These include edge distortion, spatial dependence, and broad spatial dispersion around the true hit. A fine training grid is needed to achieve approximately $10~\mu\mathrm{m}$ precision, which in turn requires large training datasets and high memory usage. In addition, the first model loses resolution under rasterisation and shaping. We, therefore, investigated a topology-agnostic model that can be applied to arbitrary pad counts and geometries, assuming otherwise consistent charge-sharing behaviour, such as compatible pad shapes and active-area depths.  The model takes pad coordinates as input, and its backbone contains spatial encoding, a transformer, and pad-coordinate weighting. In this transformer-based model, each token attends to every other token, so information flows across pads. For charge-sharing reconstruction, this is advantageous because cross-channel correlations are part of the physical response. The use of a transformer also supports an arbitrary number of tokens, making the architecture suitable for adapting to different pad counts.

Our model converts waveforms from different channels into learned feature representations, allows them to communicate with each other in the multi-head self-attention (MHSA) layer, then refines the shared information in the non-linear layer to produce a weighted sum of the pad coordinates. More precisely, the inputs consist of $c$ channels associated with waveforms of length $t$ and 2 spatial coordinates of the readout pads. A linear embedding layer projects the waveforms and the 2D pad coordinates into a shared latent dimension $d$. These embeddings are summed to form a set of $c$ input token embeddings, which are then processed by a transformer-style network with layer normalisation and residual connection. The first layer is an MHSA block that enables non-local feature sharing across temporally separated features in different channels. It works by projecting its input into query ($Q$), key ($K$), and value ($V$) tensors (the many heads allow multiple features to be learned simultaneously). $Q$ and $K$ compute self-attention maps via softmax that encode the importance of each channel relative to the others, and are applied to the values. The outputs from each attention head are concatenated and projected back into a $c$-by-$d$ representation. This is followed by a position-wise feed-forward network consisting of an up-projection, non-linear activation, and down-projection. The resulting embedding is linearly mapped into a scalar weight for each pad, from which we obtain the final predicted coordinate.

\begin{figure}[htbp]
\centering
\includegraphics[width=0.4\textwidth]{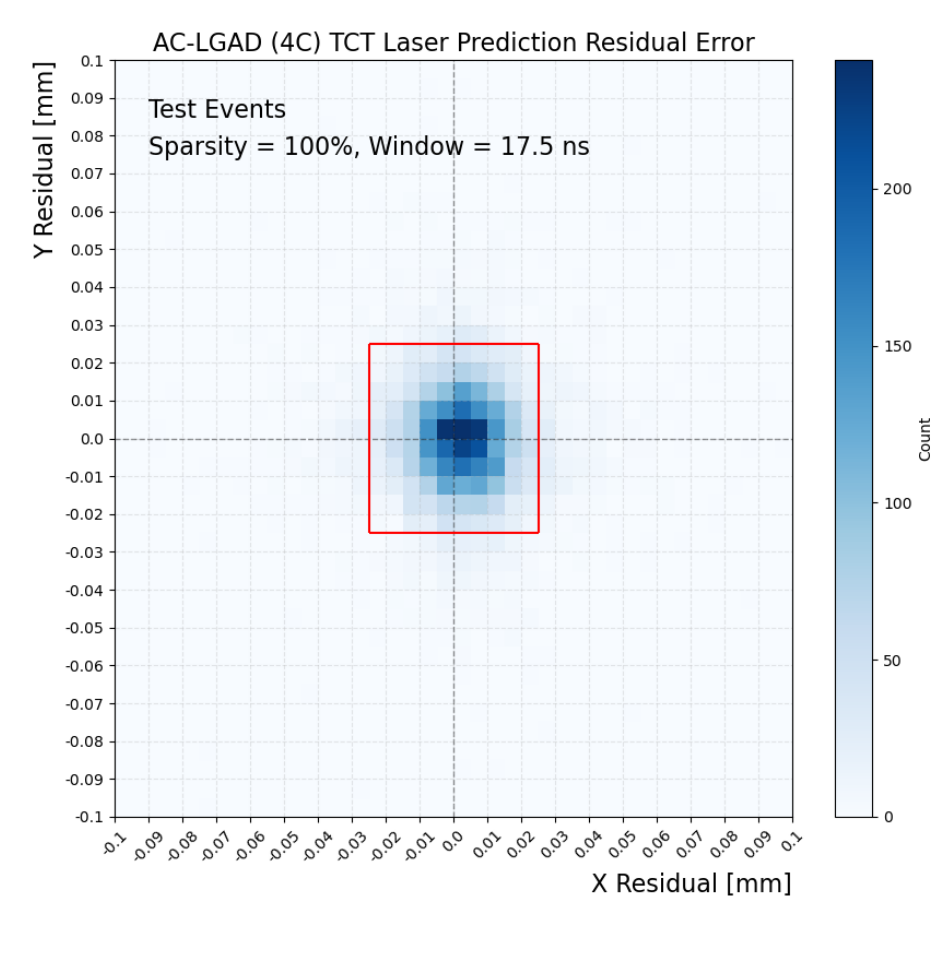}
\includegraphics[width=0.4\textwidth]{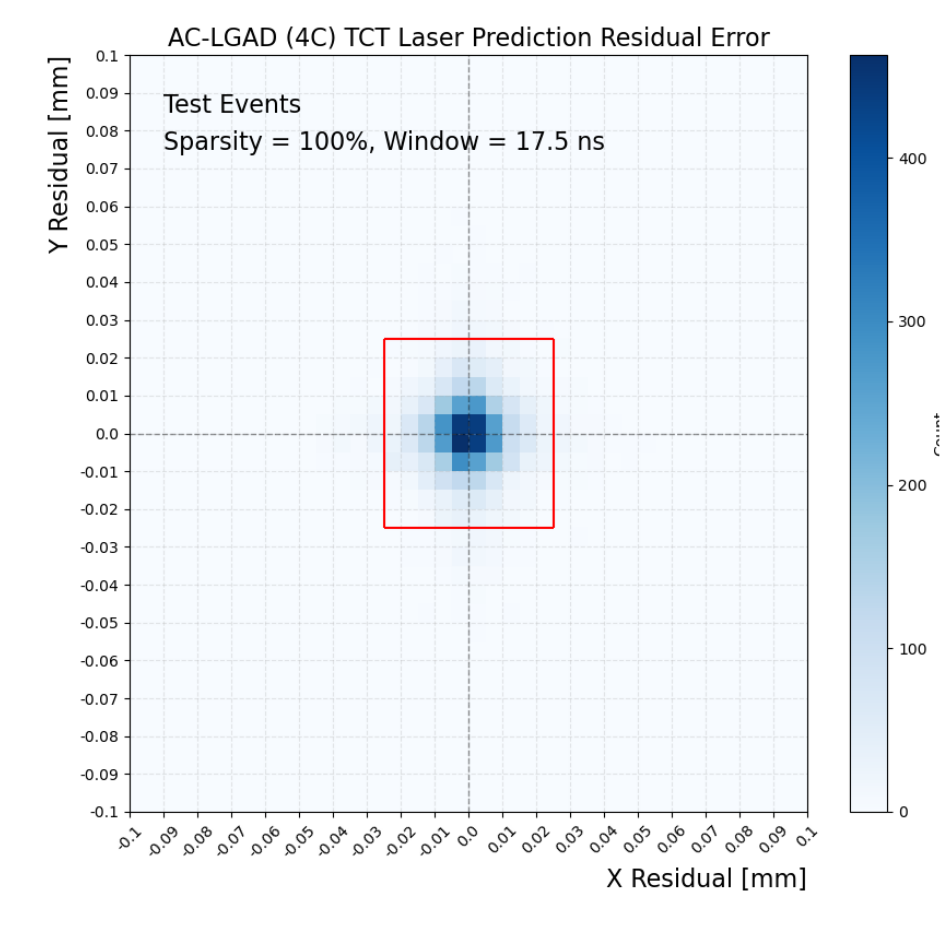}
\caption{Bidimensional hit residuals for the RNN architecture (left) and the transformer-based architecture (right). \label{fig:residual-comparison}}
\end{figure}

This transformer-based model mitigates edge distortions and yields clusters closer to the true positions. The improvement is clearest near the edges of the region of interest, where one or two pads carry most of the charge. We also observe a narrower spread around the true position. Residual distributions also improve, see Fig.~\ref{fig:residual-comparison}. The residuals are overall smaller, and less information is needed to target the same precision. 
In the comparison shown, the CIM residuals are more clustered around zero than the LSTM residuals, while the LSTM baseline performs the worst. 

\begin{figure}[htbp]
\centering
\includegraphics[width=\textwidth]{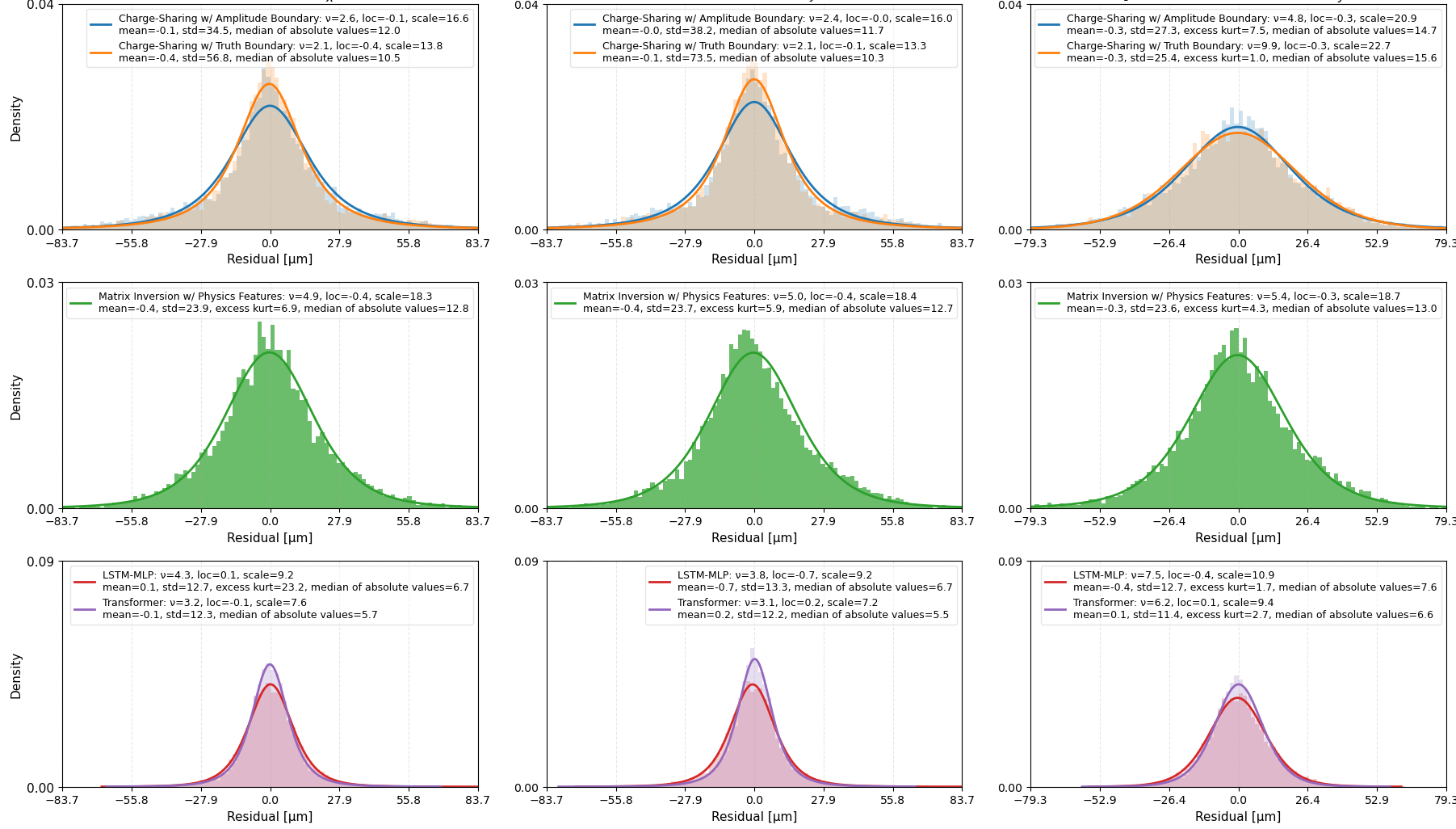}
\caption{One-dimensional hit-residual distributions fitted with Student-$t$ functions. Columns show the residual in the $x$ direction, $r_x$, the residual in the $y$ direction, $r_y$, and the projected residual $(r_x+r_y)/\sqrt{2}$. Rows compare different reconstruction approaches: charge sharing using the amplitude-defined boundary and the truth-defined boundary, matrix inversion with physics-inspired features, and the learned LSTM and transformer-based models. The histograms show the residual distributions, while the solid curves show the corresponding Student-$t$ fits. Each legend reports the fitted Student-$t$ parameters, including the degrees of freedom $\nu$, location, and scale, together with summary statistics such as the mean, standard deviation, excess kurtosis, and median absolute residual. Smaller widths and median absolute residuals indicate better spatial resolution, while lower $\nu$ and larger excess kurtosis indicate heavier non-Gaussian tails and a larger population of outlier events. \label{fig:residual-studentt}}
\end{figure}

Lastly, we also study how the shape of the residual distributions changes depending on the models considered. Since the residuals are visibly non-Gaussian, we fit the one-dimensional residual projections with Student-$t$ functions, which provide a compact description of both the central resolution and the non-Gaussian tails. The results are summarised in Fig.~\ref{fig:residual-studentt}. The charge-sharing methods exhibit broad distributions with low Student-$t$ degrees of freedom, with $\nu \simeq 2.1$ to $-2.6$ for the individual $x$ and $y$ residuals, indicating heavy tails and a sizeable population of large-error events. This is also reflected in the large residual widths, with standard deviations ranging from approximately $35$ to $57~\mu\mathrm{m}$ in $x$ and from $38$ to $74~\mu\mathrm{m}$ in $y$, and median absolute residuals of about $10$ to  $15~\mu\mathrm{m}$. Matrix inversion reduces the overall spread, giving standard deviations of about $24~\mu\mathrm{m}$ in $x$, $y$, and the projected residual, with median absolute residuals of approximately $13~\mu\mathrm{m}$. However, its fitted degrees of freedom remain moderate, $\nu \simeq 4.9$ to  $5.4$, showing that noticeable non-Gaussian tails are still present, as expected. In contrast, the LSTM and transformer-based models produce much narrower residual cores and substantially smaller absolute tail excursions. The LSTM  gives residual standard deviations of about $12.7~\mu\mathrm{m}$ in $x$, $13.3~\mu\mathrm{m}$ in $y$, and $12.7~\mu\mathrm{m}$ for the projected residual, while the transformer gives approximately $12~\mu\mathrm{m}$, and $11~\mu\mathrm{m}$, respectively. The corresponding median absolute residuals are reduced to about $6$ to  $8~\mu\mathrm{m}$, with the transformer reaching $5.7~\mu\mathrm{m}$ in $x$, $5.5~\mu\mathrm{m}$ in $y$, and $6.6~\mu\mathrm{m}$ for the projected residual. Although the full waveform regression ML-based model residuals are still better described by Student-$t$ shapes than by purely Gaussian functions, their smaller scale parameters, around $7$ to  $11~\mu\mathrm{m}$ compared with roughly $16$ to $21~\mu\mathrm{m}$ for charge sharing and $18$ to $19~\mu\mathrm{m}$ for matrix inversion, show that the tail events are significantly compressed in absolute size. This indicates that the transformer-based reconstruction improves not only the average position resolution, but also the robustness of the reconstruction against rare, poorly reconstructed hits. 

\begin{figure}[htbp]
\centering
\includegraphics[width=0.5\textwidth]{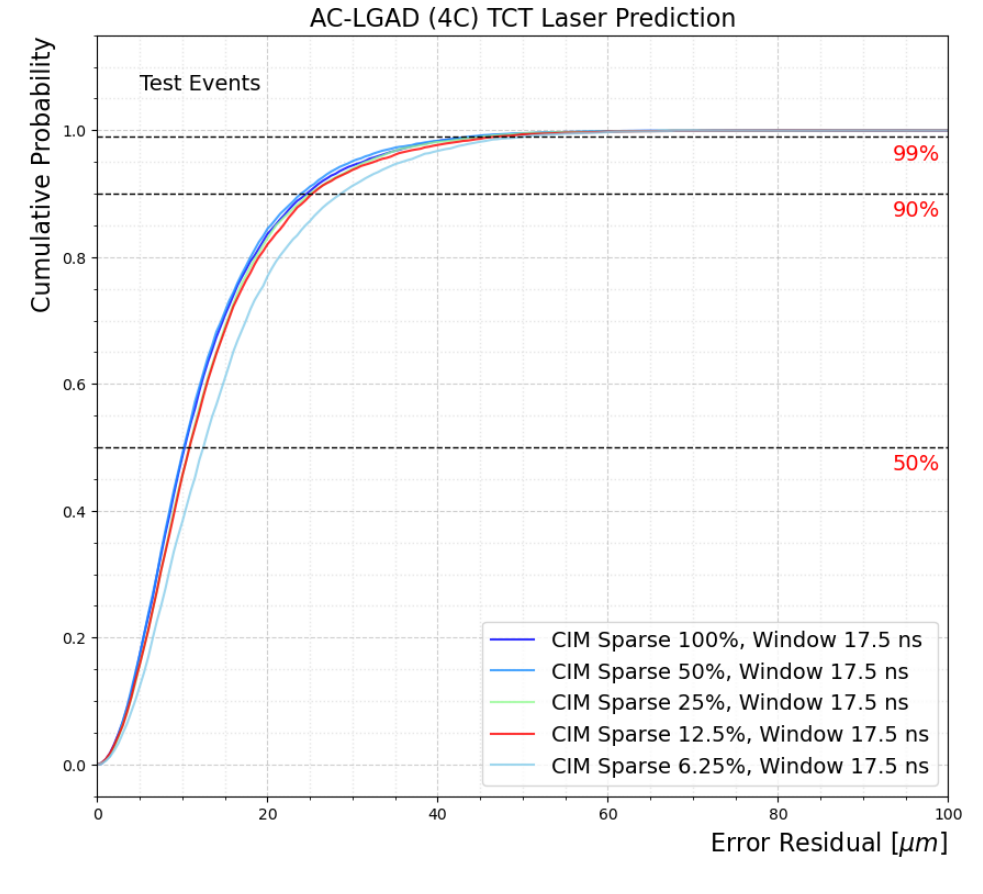}
\caption{The figure shows the cumulative hit uncertainty probability for the geometry-based model.\label{fig:sparsity-performance}}
\end{figure}

Our performance studies indicate that the model maintains a position resolution of around $10~\mu\mathrm{m}$ for $500~\mu\mathrm{m}\times500~\mu\mathrm{m}$ pitched sensors. The resolution loss with decreased sampling is significantly shallower than in previous studies. The rise in precision loss occurs at about 95\% sampling reduction, compared with about 70\% in earlier studies. The model preserves approximately $10~\mu\mathrm{m}$ position resolution for $500~\mu\mathrm{m}\times500~\mu\mathrm{m}$ pitched sensors and shows no significant performance drop until 12.5\% sparsity, as shown in Fig.~\ref{fig:sparsity-performance}.

\section{Conclusions}

Several open questions remain under investigation. With domain adaptation across different bias points, we intend to mitigate dependence on shaping and charge. Ionisation dependence will be studied by expanding transformer layers for MIP response in test beams, conditioned on large-charge TCT datasets. Geometric independence is being stressed by expanding the current model to larger datasets using sensors with varied geometries. Additional detector effects, including ballistic deficits and reflections, are also under study. An irradiation campaign in Ljubljana with the same sensors is ongoing to understand and model the post-irradiation response. Expansion into timing capabilities is foreseen for the current model, using constant fraction discriminators and position-output targets.

When normalized to the sensor pitch, the achieved resolution is competitive with published AC-LGAD and RSD reconstruction results, where position resolutions at the level of a few percent of the pitch have been reported using charge-sharing and machine-learning methods~\cite{siviero2021,investigationchargesharing,tornago2023,heller2022_aclgad_bnl_hpk_jinst,arcidiacono2023}. The main contribution of the proposed approach pertains rather to the combination of competitive pitch-normalized performance with improved rare events modelling: the residual distributions are narrower and the non-Gaussian tails are reduced with respect to the charge-sharing, matrix-inversion, and LSTM methods considered in this work. Although the comparison with other methods in the literature is not one-to-one, since the values depend strongly on sensor geometry, pitch, readout pitch and capacitance, number of readout channels, laser or beam-test conditions, and the definition of the quoted resolution, the transformer-based reconstruction reaches a projected residual width of about $10~\mu\mathrm{m}$, with a median absolute residual of about $6.6~\mu\mathrm{m}$, while using a topology-agnostic architecture. This is substantially better than the charge-sharing and matrix-inversion baselines studied here, and is comparable to the best reported AC-LGAD/RSD results.

Optimising the information contained in the charge collection of resistive silicon detectors can accelerate improved spatial resolution with current
production technologies and guide future sensor designs. AC-LGADs provide a 100\% fill factor and enable good spatial resolution with relaxed pitch through charge sharing, but the same charge sharing produces coupled multi-channel waveforms and increases readout complexity. Full-waveform machine-learning reconstruction addresses this complexity by using correlations across pads and across waveform samples. The first LSTM-based models demonstrate the value of the approach and have been tested for a possible deployment on FPGAs. Compression studies show that substantial waveform reduction is possible while preserving the relevant peak region. The newer topology-agnostic transformer-based direction reduces geometry dependence, mitigates edge distortions, and supports arbitrary pad counts and geometries. These developments point toward ML-assisted readout and design of future resistive silicon sensors.

\bibliographystyle{unsrt}
\bibliography{main.bib}

\end{document}